\documentstyle[preprint,epsf]{jpsj}
\def\runtitle{ 
Tomonaga-Luttinger-Liquid Theory of Metallic Carbon Nanotubes with Open Boundaries 
}
\def\runauthor{Hideo {\sc Yoshioka} and Yoko {\sc Okamura}}
\def\d{{\rm d}}

\def\e{{\rm e}}

\def\ggs{\buildrel\textstyle > \over {\hbox{\raise0.2ex\hbox{$\sim$}}}}
\def\lls{\buildrel\textstyle < \over {\hbox{\raise0.2ex\hbox{$\sim$}}}}
\def\gsim{\,\lower0.75ex\hbox{$\ggs$}\,}
\def\lsim{\,\lower0.75ex\hbox{$\lls$}\,}

\def\im{{\rm i}}
\def\ie{{\it i.e.}, }
\def\delx{\partial_x}
\def\al{\alpha}

\def\bvec #1{\mbox{\boldmath $#1$}}
\def\tia{{\tilde a}}
\def\jo #1#2#3#4{#1 {\bf #2} (#3) #4}   

\def\PRB{Phys.\ Rev.\ B}
\def\PRL{Phys.\ Rev.\ Lett.}

\def\APL{Appl.\ Phys.\ Lett.}

\def\JPC{J.\ Phys.\ C}

\def\EPL{Europhys.\ Lett}

\def\NAT{Nature (London)}
\def\EPJB{Eur.\ Phys.\ J.\ B}

\title
{
Tomonaga-Luttinger-Liquid Theory of Metallic Carbon Nanotubes with Open Boundaries 
}

\author{
Hideo {\sc Yoshioka}\hspace{-0.5mm}\footnote{E-mail:
yoshioka@phys.nara-wu.ac.jp}\hspace{0.5mm} 
and Yoko {\sc Okamura}\hspace{-0.5mm}\footnote{E-mail:
yoko@phys.nara-wu.ac.jp}\hspace{0.5mm}
}

\inst{
Department of Physics, Nara Women's University, Nara 630-8506
}

\recdate{May 20, 2002}

\abst{
Tomonaga-Luttinger-liquid theory is formulated for metallic 
carbon nanotubes with open boundaries.
Both cases of single- and multi-wall nanotubes are discussed. 
Based on this theory, 
spatial variation of the charge density from an edge is investigated 
with taking account of the shift of the chemical potential 
which expresses the carrier injection to the nanotube. 
The charge density has the spatially independent part and the
oscillatory component.  
Roles of Coulomb interaction on the amplitude 
of the oscillation, the wavenumbers of it and the uniform component of the
charge density are clarified. 
}

\kword{carbon nanotubes, Tomonaga-Luttinger-liquid, bosonization, Friedel oscillation
}

\begin{document}
\sloppy
\maketitle


\section{Introduction}

A carbon nanotube
is composed of a coaxially rolled 
graphite sheet\cite{Iijima} 
and it's actual length is of the order of $1\mu m$ or less than it. 
The material is characterized by two integers, $(N_+, N_-)$, 
corresponding to a wrapping vector along the waist, 
$\bvec w = N_+ \bvec a_+ + N_- \bvec a_-$, 
where $\bvec a_{\pm} = (\pm a/2, \sqrt{3}a/2)$ 
are primitive lattice vectors of 
the graphite and $|\bvec a_{\pm}| = a$ is the lattice spacing.
It has been shown that the carbon nanotubes have peculiar 
band structures.\cite{Hamada,Saito-I,Saito-II} 
When $N_+ - N_- = 0$ mod 3, 
the metallic one-dimensional 
dispersions appear near the center of the bands. 
The low energy properties less than $v_0 / R$ 
($v_0$ : Fermi velocity, $R$ : radius of the tube) are 
well described by taking into account only the metallic one-dimensional
dispersions.
Thus, the metallic carbon nanotubes are considered 
as the typical one-dimensional conductors. 

It has been well known that
physical properties of the one-dimensional interacting electron
systems cannot be described by conventional Fermi-liquid-theory. 
Instead, the systems show the behaviors called as 
Tomonaga-Luttinger-liquid. 
The Tomonaga-Luttinger-liquid state is characterized by 
separation of the charge and spin degrees of freedom, and 
the anomalous exponents of correlation functions, 
which depend on the interaction.   
Carbon nanotubes are one of the most promising candidates 
where such exotic correlation effects can be observed. 
The electronic states of the metallic carbon nanotube
have been theoretically investigated with taking account of 
the long-range Coulomb interaction\cite{EG-I,Kane,EG-II,YO,OY,OSY} and  
the novel correlation effects have been found.   
Experimentally, in the transport measurements for single-wall nanotubes
(SWNTs)\cite{Bockrath,Yao,Postma},  
power-law dependences of the conductance as a function of temperature
and of the differential conductance as a function of bias voltage have
been observed in the metal-SWNT  junctions and in the SWNT-SWNT junctions. 
These results have been interpreted to be due to tunneling between 
the Fermi-liquid and the Tomonaga-Luttinger-liquid in the former case, 
and that between 
the Tomonaga-Luttinger-liquid and the Tomonaga-Luttinger-liquid in the latter case. 

The correlation effects of the semi-infinite\cite{Kane,EG-II} and 
finite length carbon nanotube\cite{Kane} have been
studied theoretically. 
The local density of state of the semi-infinite SWNT calculated by the
bosonization theory explains the transport experiments 
quantitatively. 
In ref.\citen{Kane},  
the local tunneling density of states of the finite length carbon nanotube 
has been investigated at absolute zero temperature ($T=0$).  
It has been found that the new energy scale, which reflects the
spin-charge separation in the Tomonaga-Luttinger-liquid, 
appears in Coulomb blockade behavior in addition to usual charging energy and
single-particle level spacing. 
However, in the above theories, 
the relationship between the original electron operator and 
the slowly varying Fermi field describing the low energy physics 
has not been clarified.  
The relationship is important for discussing 
the concrete physical quantities, especially the spatial dependence of those
with rapid oscillation determined by Fermi wavelength. 
Note that the oscillatory component of the local density of states 
has been neglected in refs. \citen{Kane} and \citen{EG-II} 
because the transport measurements are considered to observe the local
density of states averaged over several lattice constants.     
In the present paper, we develop the bosonization theory 
of the metallic carbon nanotube with open boundaries based on the theory
for one-dimensional system\cite{Fabrizio} 
with paying attention to the relationship between
the two kinds of electron operators. 
The theory is extended to the case of the multi-wall nanotube (MWNT). 
By utilizing the theory, 
the spatial variation of the charge density from an edge 
is investigated in the presence of the shift of
the chemical potential expressing doping of carriers to the nanotube.   
It is found that the uniform component of the charge density and  
the wavenumber of the oscillation become smaller compared with the
non-interacting case due to the effects of the interaction on the zero modes.     
The amplitude of the oscillation determined by the bosonic fluctuation 
is shown to be larger than that in the
absence of the interaction. 
The amplitude for MWNT depends on the number of the metallic shells
included in the MWNT, and the effect of the interaction on the
amplitude vanishes when the number tends to infinity.    

The organization of this paper is as follows. 
In \S.2, the bosonization theory with open boundaries is developed for
the SWNT. 
The theory is extended to the case of the MWNT in \S.3. 
The charge distribution of the SWNT and of the MWNT are calculated in
\S.4. 
Section 5 is devoted to summary and discussion.

\section{Bosonization for SWNTs with open boundaries}
We formulate the bosonization theory 
of the metallic SWNT with open boundaries. 
As a model of the metallic carbon nanotube, we consider the $(N,N)$ armchair
nanotube.   
\subsection{Non-interacting case}
We consider the armchair carbon nanotube with the length, $L$, and 
the radius, $R = \sqrt{3}Na/(2\pi)$, schematically shown in Fig. 1. 
The Hamiltonian in the tight-binding model is written as  
${\cal H} = {\cal H}_{\rm k} + {\cal H}_{\rm int}$, 
where ${\cal H}_{\rm k}$ is the kinetic part,   
\begin{eqnarray}
{\cal H}_{\rm k} &=& -t \sum_{{\bvec r}, s, p} a^\dagger_{p,s}({\bvec r})
\big\{
a_{-p,s}({\bvec r}) 
+ a_{-p,s}({\bvec r}-p{\bvec a}_+) 
+ a_{-p,s}({\bvec r}-p{\bvec a}_-)
\big\},  
\label{eqn:Hk}
\end{eqnarray}
and ${\cal H}_{\rm int}$ describes the mutual interaction,
\begin{eqnarray}
{\cal H}_{\rm int} &=& \frac{1}{2} \sum_{{{\bvec r},{\bvec r}'}}
 \sum_{s, s'} \sum_{p,p'} U({\bvec r} - {\bvec r}' - (p-p'){\bvec d}/2) 
a^\dagger_{p,s}({\bvec r}) a^\dagger_{p',s'}({\bvec r}')
a_{p',s'}({\bvec r}') a_{p,s}({\bvec r}). 
\label{eqn:Hint}
\end{eqnarray}
Here
$t$ denotes the hopping integral between the nearest-neighbor atoms, 
and $a^\dagger_{p,s} ({\bvec r})$
is the creation operator of the electron with spin $s = \pm$ 
at the location ${\bvec r} - p {\bvec d}/2$ where 
$\bvec r = (a l, \sqrt{3}a m)$ or $(a (l+1/2), \sqrt{3}a(m+1/2))$ 
($l$, $m$ : integer), $p=\pm$ 
and $\bvec d = (0,a/\sqrt{3})$. 
The interaction, $U({\bvec r})$, is given by 
$U({\bvec r}) = e^2/\left\{\kappa \sqrt{a_0^2 + x^2 + 4R^2
\sin^2(y/2R)}\right\}$\cite{EG-I}  with ${\bvec r} = (x,y)$
where $e$, $\kappa$ and  $a_0$ are 
the electric charge, dielectric constant and short-range cut-off of the
interaction,  respectively. 
  \begin{figure}
\vspace{2em}
\centerline{\epsfxsize=7.0cm\epsfbox{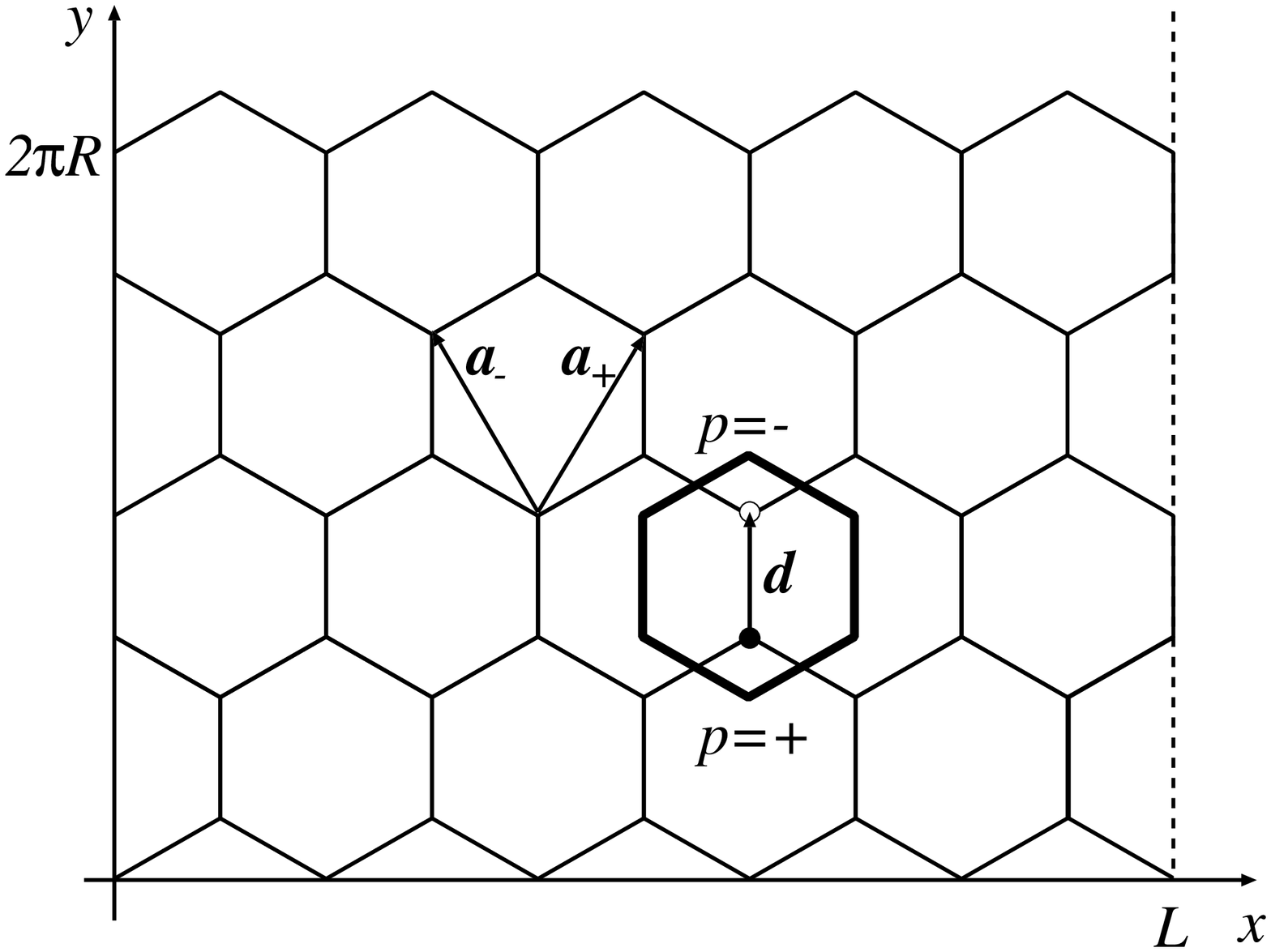}}
\caption{ Carbon atoms in the armchair nanotubes with length, $L$, and 
the radius, $R$,  
where the $x$($y$) axis is along the tube (waist). 
Here $\bvec a_{\pm}$ are two primitive lattice vectors of graphite, 
$|\bvec a_{\pm}| = a$, and ${\bvec d} = (0,a/\sqrt{3})$.   
The hexagon delineated by the thick line is the unit cell and 
the black (white) circle denotes the sublattice $p=+(-)$.
} 
    \label{fig:CN}
    \end{figure}
  \begin{figure}
\vspace{2em}
\centerline{\epsfxsize=7.0cm\epsfbox{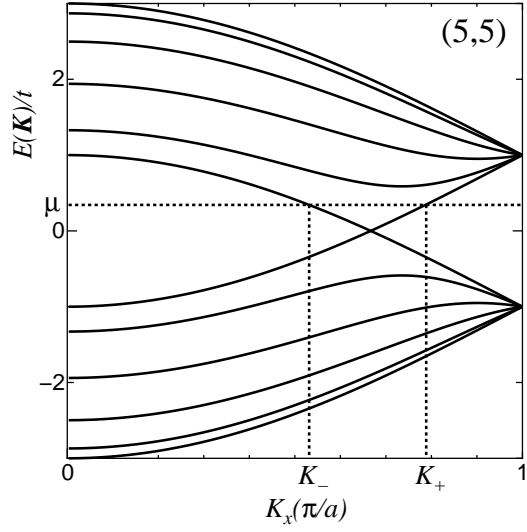}}
\caption{ Energy dispersion, $E({\bvec K})$, of the (5,5) armchair nanotube as
a function of $K_x$.
Fermi energy in the absence of the chemical potential, $\mu$, is $E({\bvec K}) = 0$
and $K_{\pm} \simeq K_0 \pm \mu/v_0$ where $K_0 = 2 \pi / (3 a)$ and
   $v_0 = \sqrt{3}t a / 2$.  
} 
    \label{fig:Band}
    \end{figure}

At first, we consider the non-interacting case. 
In order to diagonalize ${\cal H}_{\rm k}$,  
we solve Schr$\ddot{\rm o}$dinger equation, 
\begin{eqnarray}
& & -t \left\{\phi_{{\bvec K}}^{r,-p} ({\bvec r}) 
+ \phi_{{\bvec K}}^{r,-p} ({\bvec r} -p {\bvec a}_+)
+ \phi_{{\bvec K}}^{r,-p} ({\bvec r} -p {\bvec a}_-)\right\} 
 = E^r(\bvec K) \phi_{{\bvec K}}^{r,p} ({\bvec r}),   
\label{eqn:S-eq}
\end{eqnarray}
with the boundary condition, 
$\phi_{{\bvec K}}^{r,p} (0,y) = \phi_{{\bvec K}}^{r,p} (L,y) = 0$
and $\phi_{{\bvec K}}^{r,p} ({\bvec r}) 
= \phi_{{\bvec K}}^{r,p} ({\bvec r} + {\bvec w})$  
where ${\bvec w} = N ({\bvec a}_+ + {\bvec a}_-) = (0,\sqrt{3}Na)$.
The solution and eigenvalue are obtained as follows, 
\begin{eqnarray}
 \phi_{{\bvec K}}^{r,p} ({\bvec r}) &=& 
\sqrt{\frac{a}{NL}} A^{r,p}({\bvec K}) \sin K_x x \e^{\im K_y y}, \\
E^r({\bvec K}) &=& -r t |\alpha({\bvec K})|, 
\end{eqnarray} 
where $r = \pm$, $A^{r,+}({\bvec K}) = 1/\sqrt{2}$ and 
$A^{r,-}({\bvec K}) = r \alpha ({\bvec K})/(\sqrt{2} |\alpha ({\bvec
K})|)$
with $\alpha ({\bvec K}) = 1 + 2 \cos (K_x a/2) \e^{\im K_y \sqrt{3}
a/2}$.
The wavenumbers, $K_x$ and $K_y$, are quantized as  
$K_x = n_x \pi/L$ $(n_x = 1, 2, \cdots, L/a)$
and $K_y = 2 \pi n_y/(\sqrt{3}Na)$ $(n_y = 1, 2, \cdots, 2N)$.
We note that the energy dispersions with 
$K_y = 2 \pi /(\sqrt{3}a) \equiv K_y^0$ 
touch the Fermi energy ($E^r(\bvec K) = 0$) 
at $K_x = 2 \pi /(3 a) \equiv K_0$, and the others 
do not cross the Fermi energy.  
The Hamiltonian is written as 
${\cal H}_{\rm k} = \sum_{\bvec K} \sum_{r,s} E_r({\bvec K}) 
c^\dagger_{r,s}({\bvec K})c_{r,s}({\bvec K})$
where $a_{p,s}({\bvec r}) = \sum_{\bvec K} \sum_{r} \phi_{{\bvec
K}}^{r,p} ({\bvec r}) c_{r,s}({\bvec K})$.  
When the chemical potential, $\mu$, stays the energy bands with 
$K_y = K_y^0$ (see Fig.2),  
the charge density in the absence of the interaction 
is given at $T=0$ for the semi-infinite case, $L \to \infty$, as 
\begin{eqnarray}
& &\langle:\sum_{s} a^\dagger_{p,s}({\bvec r}) a_{p,s}({\bvec r}):\rangle 
= \frac{(K_+ - K_-)a}{2 N \pi}
- \frac{a\left(\sin 2 K_+ x - \sin 2K_-x \right)}{4N\pi x}, 
\end{eqnarray} 
where $K_\pm \simeq K_0 \pm \mu/v_0$ and 
$v_0 = \sqrt{3}ta/2$. 
The charge distribution has two parts. 
One is the uniform part due to the shift of the chemical potential 
and the other expresses the oscillation (Friedel oscillation). 
We find that there exit the two kinds of oscillation with the wavenumbers, $2K_\pm$, 
and those are out of phase. 
The oscillatory behavior is absent for $\mu = 0$.  

We apply the bosonization procedure 
on the system with the open boundaries\cite{Fabrizio} to the present model.
Energy dispersion is linearized by substituting ${\bvec K} = (K_0 + k,
K_y^0)$ into $E^r(\bvec K)$. 
When we define the right/left moving electron operator, 
$a_{R/L,s}(K_0+k)$, as
\begin{eqnarray}
 a_{R,s}(K_0 + k) &=& 
\left\{
\begin{array}{ll}
 c_{-,s}({\bvec K}), & \quad k \geq 0 \\
 c_{+,s}({\bvec K}), & \quad k < 0 \\
\end{array}
\right. \\ 
 a_{L,s}(K_0 + k) &=& 
\left\{
\begin{array}{ll}
 c_{+,s}({\bvec K}), & \quad k \geq 0 \\
 c_{-,s}({\bvec K}), & \quad k < 0, \\
\end{array}
\right.
\end{eqnarray}
the Hamiltonian, ${\cal H}_{\rm k}$, 
and the electron operator
are respectively written as follows,
\begin{eqnarray}
{\cal H}_{\rm k} &=& \sum_{k,s} 
v_0 k \left\{ a^\dagger_{R,s}(K_0 + k)a_{R,s}(K_0 + k) 
 - a^\dagger_{L,s}(K_0 + k)a_{L,s}(K_0 + k) \right\}, \\
a_{p,s}({\bvec r}) &=& \sqrt{\frac{a}{2NL}} 
\sum_{k} \sin(K_0 + k)x \e^{\im K_y^0 y} 
\left\{p a_{R,s}(K_0 + k) + a_{L,s}(K_0 + k)\right\}.  
\end{eqnarray} 
Here we define the slowly varying Fermi field, $\psi_{R/L,\alpha,s}(x)$, 
as
\begin{eqnarray}
 \psi_{R,+,s}(x) &=& \frac{-\im}{\sqrt{2L}} \sum_{k} \e^{\im k x} a_{R,s}(K_0 + k), \\
 \psi_{L,+,s}(x) &=& \frac{-\im}{\sqrt{2L}} \sum_{k} \e^{\im k x} a_{L,s}(K_0 + k), \\
 \psi_{L,-,s}(x) &=& \frac{\im}{\sqrt{2L}} \sum_{k} \e^{-\im k x} a_{R,s}(K_0 + k), \\
 \psi_{R,-,s}(x) &=& \frac{\im}{\sqrt{2L}} \sum_{k} \e^{-\im k x} a_{L,s}(K_0 + k), 
\end{eqnarray}
where $\alpha = \pm$ is the index corresponding to valleys for the case
for periodic boundary condition. 
The electron operator is expressed by the slowly varying fields as 
\begin{eqnarray}
 a_{p,s}(\bvec r) &=& \sqrt{\frac{a}{4N}}\e^{\im K_y^0 y} \left\{
\e^{\im K_0 x}\left[p \psi_{R,+,s} (x) + \psi_{L,+,s} (x)\right]
+ \e^{-\im K_0 x}\left[p \psi_{L,-,s} (x) + \psi_{R,-,s} (x)\right] \right\}.
\end{eqnarray}
The expression is the same as that in case of the periodic boundary
condition. 
The slowly varying  field operators are not independent, but satisfy,
\begin{eqnarray}
 \psi_{L,+,s}(x) &=& - \psi_{R,-,s}(-x), \\ 
  \psi_{L,-,s}(x) &=& - \psi_{R,+,s}(-x).
\end{eqnarray} 
Therefore we can actually work with the right moving operator only. 
The boundary condition, $a_{p,s}(0,y) = 0$, 
is automatically satisfied due to eqs. (16) and (17). 
However, the condition, $a_{p,s}(L,y) = 0$, implies, 
\begin{eqnarray}
 \psi_{R,\al,s}(-L) &=& \e^{\im 2 \al K_0 L}\psi_{R,\al,s}(L). 
\end{eqnarray}   
So we can regard the field $\psi_{R,\al,s}(x)$ as defined for $0 \leq x
\leq 2L$ with the boundary condition,
\begin{eqnarray}
 \psi_{R,\al,s}(x+2L) &=& \e^{-\im 2 \al K_0 L}\psi_{R,\al,s}(x) \nonumber \\
&=& \e^{\im 2 \al \pi \nu / 3}\psi_{R,\al,s}(x),   
\end{eqnarray} 
where $\nu = 0, \pm 1$ is defined as $L = (3 n_L + \nu)a$ ($n_L$ : integer
satisfying $L > 0$).
In terms of the right moving operators, 
eqs.(9) and (10) are expressed as 
\begin{eqnarray}
 {\cal H}_{\rm k} &=& v_0 \sum_{\alpha, s} \int_{-L}^{L} \d x
\psi_{R,\alpha,s}^\dagger (-\im \delx) \psi_{R,\alpha,s}, \\
 a_{p,s}({\bvec r}) &=& \e^{\im K_y^0y} \sqrt{\frac{a}{4N}}
\sum_{\alpha} p^{(1+\alpha)/2} 
\left\{
\e^{\im \alpha K_0 x} \psi_{R,\alpha,s}(x)
- \e^{-\im \alpha K_0 x} \psi_{R,\alpha,s}(-x)
\right\}. 
\end{eqnarray} 
The above Hamiltonian and the right moving electron operators
are straightforwardly bosonized by utilizing Haldane prescription.\cite{Haldane}
The bosonized form for $\psi_{R,\al,s}$ is given by 
\begin{eqnarray}
\psi_{R,\al,s}(x) &=& \frac{\eta_{\al,s}}{\sqrt{2 \pi \tia}}\e^{-\im \theta_{\al,s}}
\e^{\im \frac{\pi x}{L}\left\{\Delta N_{\al,s} + \al \nu /3\right\}}
\e^{\im \phi_{\al,s}(x)},  
\end{eqnarray}
where $\Delta N_{\al,s}$ is the extra electron with the index $(\alpha,s)$
and satisfies, $[\theta_{\al,s}, \Delta N_{\al',s'}] = \im
\delta_{\al,\al'}\delta_{s,s'}$.
The operator, $\eta_{\al,s}$, is Majorana Fermion satisfying 
$\{\eta_{\al,s},\eta_{\al',s'}\} = 2 \delta_{\al,\al'}\delta_{s,s'}$ 
and $\tia$ is the ultra violet cut-off of the order of $1/R$.  
The function, $\phi_{\al,s}(x)$, is given by 
\begin{eqnarray}
 \phi_{\al,s}(x) &=& \sum_{q > 0} \sqrt{\frac{\pi}{q L}}
\left\{
\e^{\im q x - \tia q/2} b_{\al,s} (q) + {\rm h.c.}
\right\},
\end{eqnarray}
and satisfy $\phi_{\al,s}(x+2L) = \phi_{\al,s}(x)$ 
because $q = n \pi /L$ ($n$ : integer).    
Here $b_{\al,s}(q)$ is the boson operator, so satisfies  
$[b_{\al,s}(q),b^\dagger_{\al',s'}(q')] =
\delta_{\al,\al'}\delta_{s,s'}\delta_{q,q'}$. 
The density operator, $\rho_{R,\al,s}(x) \equiv \psi^\dagger_{R,\al,s}(x)\psi_{R,\al,s}(x)$, is given by 
\begin{eqnarray}
\rho_{R,\al,s}(x) &=& \frac{\Delta N_{\al,s}+\al \nu /3}{2L} + \frac{1}{2\pi} \delx \phi_{\al,s}(x),      
\end{eqnarray}
and $\rho_{L,\al,s} (x) = \rho_{R,-\al,s}(-x)$. 
The bosonized form of ${\cal H}_{\rm k}$ is given as 
\begin{eqnarray}
 {\cal H}_{\rm k} &=& \pi v_0 \sum_{\al,s} \int_{-L}^L \d x : \rho_{R,\al,s}(x) \rho_{R,\al,s}(x) : \nonumber \\ 
&=& \frac{\pi v_0}{2L} \sum_{\al,s} (\Delta N_{\al,s} + \frac{\al \nu}{3})^2 
+ \sum_{\al,s} \sum_{q>0} v_0 q b^\dagger_{\al,s}(q) b_{\al,s}(q). 
\end{eqnarray}


Before including the mutual interaction, we change the variables of
$(\al,s)$ to those for charge/spin ($j=\rho$/$\sigma$) degree of freedom
of symmetric/antisymmetric ($\delta=+/-$) combination 
between the valley index. 
We define 
\begin{eqnarray}
 b_{\rho +}(q) &=& \frac{1}{2} \sum_{\al,s} b_{\al,s}(q), \\
 b_{\rho -}(q) &=& \frac{1}{2} \sum_{\al,s} \al b_{\al,s}(q), \\
 b_{\sigma +}(q) &=& \frac{1}{2} \sum_{\al,s} s b_{\al,s}(q), \\
 b_{\sigma -}(q) &=& \frac{1}{2} \sum_{\al,s} \al s b_{\al,s}(q), \\
 \Delta N_{\rho +} &=& \sum_{\al,s} \Delta N_{\al,s}, \\
 \Delta N_{\rho -} &=& \sum_{\al,s} \al \Delta N_{\al,s}, \\
 \Delta N_{\sigma +} &=& \sum_{\al,s} s \Delta N_{\al,s}, \\
 \Delta N_{\sigma -} &=& \sum_{\al,s} \al s \Delta N_{\al,s}.  
\end{eqnarray}
 Then $[b_{j \delta}(q),b^\dagger_{j' \delta'}(q')] =
 \delta_{j,j'}\delta_{\delta,\delta'}\delta_{q,q'}$, 
$[\theta_{\al,s}, \Delta N_{\rho +}] = \im$, 
$[\theta_{\al,s}, \Delta N_{\rho -}] = \im \al$, 
$[\theta_{\al,s}, \Delta N_{\sigma +}] = \im s$ and 
$[\theta_{\al,s}, \Delta N_{\sigma -}] = \im \alpha s$.
Since the eigenvalue of $\Delta N_{\al,s}$ is an integer, 
that of $\Delta N_{j \delta}$, $Q_{j \delta}$ has a constraint, 
\begin{eqnarray}
 Q_{\rho +} + \al Q_{\rho -} + s Q_{\sigma +} + \al s Q_{\sigma -} = 4 \times \mbox{integer}.  
\end{eqnarray}
In terms of the new variables, ${\cal H}_{\rm k}$ and
$\psi_{R,\al,s}(x)$
are respectively expressed as follows, 
\begin{eqnarray}
 {\cal H}_{\rm k} &=& \frac{\pi v_0}{8 L}
\sum_{j,\delta} 
(\Delta N_{j \delta} + \frac{4 \nu}{3}\delta_{j,\rho}\delta_{\delta,-})^2 
+ \sum_{j,\delta} \sum_{q > 0} v_0 q 
b^\dagger_{j \delta}(q) b_{j \delta}(q) \nonumber \\
&\equiv& \sum_{j,\delta} {\cal H}^0_{j,\delta} \\ 
\psi_{R,\alpha,s}(x) &=& \frac{\eta_{\alpha,s}}{\sqrt{2 \pi \tia}}
\e^{- \im \theta_{\alpha, s}} 
\e^{\im \frac{\pi x}{4 L}\left\{\Delta N_{\rho +}+ \al (\Delta N_{\rho -} + 4 \nu / 3) + s \Delta N_{\sigma +} + \alpha s \Delta N_{\sigma -}\right\}}
\nonumber \\
& & \times \e^{\frac{\im}{2}\left\{\phi_{\rho +}(x) + \alpha \phi_{\rho -}(x) 
+ s \phi_{\sigma +}(x) + \alpha s \phi_{\sigma -}(x)\right\}}, 
\end{eqnarray}
with 
\begin{eqnarray}
 \phi_{j \delta}(x) &=& \sum_{q > 0} \sqrt{\frac{\pi}{q L}}
\left\{
\e^{\im q x - \tia q/2} b_{j \delta} (q) + {\rm h.c.}
\right\}. 
\end{eqnarray}

\subsection{Effects of interaction}
We take into account of the mutual interaction, eq.(2).  
Among the various interaction processes, 
the term with the strongest amplitude 
is written in terms of slowly varying Fermi fields
as,\cite{Kane,EG-I}   
\begin{eqnarray}
 {\cal H}_{\rm int}
&=& \frac{V(0)}{2}\int_{0}^L \d x
\left\{
\sum_{\alpha,s} \left[\rho_{R,\al,s}(x) + \rho_{L,\al,s}(x)\right]
\right\}^2, 
\nonumber \\ 
&=& \frac{V(0)}{2}\int_{-L}^L \d x
\left\{
\left[\sum_{\alpha,s} \rho_{R,\al,s}(x)\right]^2 
+ \sum_{\alpha,s} \sum_{\alpha',s'}\rho_{R,\al,s}(x) \rho_{R,\al',s'}(-x)
\right\},  
\end{eqnarray}
where $V(0) = (2 e^2/\kappa) \ln (R_s/R)$ where $R_s \sim L$ characterizes 
the large distance cut-off of the Coulomb interaction. 
The other interaction processes, 
whose couplings scale as $a/(2\pi R)$ and are extremely small, have been
known to play crucial roles in the absence of the carrier doping, \ie at
half-filling.\cite{Kane,YO,OY}
Since we discuss the case for away from half-filling by introducing 
the shift of the chemical potential,
those are safely neglected.   
The above interaction has a following bosonized form, 
\begin{eqnarray}
 {\cal H}_{\rm int} &=& \frac{V(0)}{2}
\left\{
\frac{1}{L}(\Delta N_{\rho +})^2 + \sum_{q>0} \frac{4q}{\pi}
b^\dagger_{\rho +}(q) b_{\rho +}(q) 
-   \sum_{q>0} \frac{2q}{\pi}
\left(
b_{\rho +}(q)b_{\rho +}(q) + b_{\rho +}^\dagger(q) b_{\rho +}^\dagger(q)
\right) 
\right\}. \nonumber \\
& & 
\end{eqnarray}
Since the interaction term, eq.(39), is expressed by $(\rho +)$ mode
only, 
the Hamiltonian except $(\rho +)$ have diagonalized form, 
${\cal H}_{j \delta}^0$.    
The Hamiltonian of $(\rho +)$ mode, ${\cal H}_{\rho +}$, is diagonalized 
by Bogoliubov transformation, 
$b_{\rho +}(q) \to \cosh \varphi b_{\rho +}(q) 
- \sinh \varphi b^\dagger_{\rho +}(q)$ with 
$\e^{2\varphi} = 1/\sqrt{1 + 4 V(0)/(\pi v_0)} \equiv
K_{\rho+}$. 
As a result, 
${\cal H}_{\rho +}$ are given as 
\begin{eqnarray}
 {\cal H}_{\rho +} &=& \frac{\pi v_{\rho + N}}{8 L} (\Delta N_{\rho +})^2 
+ \sum_{q>0} v_{\rho +} q b^\dagger_{\rho +}(q) b_{\rho +}(q), 
\end{eqnarray}
where 
$v_{\rho +} = v_0 / K_{\rho +}$ and $v_{\rho + N} = v_0 / K^2_{\rho +}$.
The quantity, $\phi_{\rho +}(x)$, is transformed as 
\begin{eqnarray}
\phi_{\rho +}(x) \to 
\cosh \varphi \phi_{\rho +}(x) - \sinh \varphi \phi_{\rho +}(-x), 
\end{eqnarray}
in eqs.(36). 
From eq.(40), the charge susceptibility per unit length 
is easily derived as $4/(\pi v_{\rho + N})$. 
The quantity is suppressed by the long-range Coulomb interaction
because $K_{\rho +} < 1$.  
For typical metallic nanotubes, the value of $K_{\rho +}$ is estimated as about
0.2,\cite{Kane}  
which leads to strong suppression of the charge susceptibility as $4/(\pi v_0)
\times 0.04$.

\section{Extension to multi-wall nanotubes}
In this section, we extend the bosonization theory with open boundaries  
on the SWNT developed in \S.2 to the case of MWNT.\cite{Egger}
We consider the MWNT where $N_M$ metallic graphite
shells with radii $R_1 < R_2 < \cdots < R_{N_M}$ are included.
For simplicity, we consider the case where all the metallic shells
consist of armchair nanotubes.   
The insulating shells in the MWNT can be incorporated in
space-dependent dielectric constant. 
The electron operator for the $n$-th shell, $a_{p,s,n}(\bvec r)$, 
is given as follows,  
\begin{eqnarray}
 a_{p,s,n}({\bvec r}) &=& \e^{\im K_y^0y} \sqrt{\frac{a}{4N_n}}
\sum_{\alpha} p^{(1+\alpha)/2} 
\left\{
\e^{\im \alpha K_0 x} \psi_{R,\alpha,s,n}(x)
- \e^{-\im \alpha K_0 x} \psi_{R,\alpha,s,n}(-x)
\right\}, 
\end{eqnarray}
where $2 \pi R_n = \sqrt{3} N_n a$ and
\begin{eqnarray}
\psi_{R,\alpha,s,n}(x) &=& \frac{\eta_{\alpha,s,n}}{\sqrt{2 \pi \tia}}
\e^{- \im \theta_{\alpha, s, n}} 
\e^{\im \frac{\pi x}{4 L}\left\{\Delta N_{\rho + n}+ \al (\Delta N_{\rho - n} + 4 \nu / 3) + s \Delta N_{\sigma + n} + \alpha s \Delta N_{\sigma - n}\right\}}
\nonumber \\
& & \times 
\e^{\frac{\im}{2}\left\{\phi_{\rho + n}(x) + \alpha \phi_{\rho - n}(x) 
+ s \phi_{\sigma + n}(x) + \alpha s \phi_{\sigma - n}(x)\right\}},
\end{eqnarray}
where $\{\eta_{\al,s,n}, \eta_{\al',s'n'}\} = 2 \delta_{\al,\al'}
\delta_{s,s'} \delta_{n,n'}$, $\al \sim 1/R_{N_M}$, 
$[\theta_{\al,s,n}, \Delta N_{\rho + n'}] = \im \delta_{n,n'}$, 
$[\theta_{\al,s,n}, \Delta N_{\rho - n'}] = \im \al \delta_{n,n'}$, 
$[\theta_{\al,s,n}, \Delta N_{\sigma + n'}] = \im s \delta_{n,n'}$,
$[\theta_{\al,s,n}, \Delta N_{\sigma - n'}] = \im \alpha s \delta_{n,n'}$,
 and  
\begin{eqnarray}
 \phi_{j \delta n}(x) &=& \sum_{q>0} \sqrt{\frac{\pi}{q L}}\left(\e^{\im q x - \tia q/2} b_{j \delta n}(q)
+ h.c. \right), 
\end{eqnarray}
with $[b_{j \delta n}(q),b^\dagger_{j' \delta' n'}(q')] = 
\delta_{j,j'} \delta_{\delta,\delta'} \delta_{n,n'} \delta_{q,q'}$. 
The Hamiltonian is written in terms of the bosonic variables as    
\begin{eqnarray}
 {\cal H}_{\rm k} &=& \frac{\pi v_0}{8 L}
\sum_{n=1}^{N_M} \sum_{j,\delta} 
(\Delta N_{j \delta n} + \frac{4 \nu}{3}\delta_{j,\rho}\delta_{\delta,-})^2 
+ \sum_{n=1}^{N_M} \sum_{j,\delta} \sum_{q > 0} v_0 q 
b^\dagger_{j \delta n}(q) b_{j \delta n}(q),    \\
{\cal H}_{\rm int} &=& \sum_{n,m = 1}^{N_M} \frac{V_{nm}}{2} \int_{-L}^{L} \d x 
\sum_{\al,s} \sum_{\al',s'}
\left\{
\rho_{R,\al,s,n}(x) \rho_{R,\al',s',m}(x) + \rho_{R,\al,s,n}(x) \rho_{R,\al',s',m}(-x) 
\right\} \nonumber \\
&=& \frac{\pi v_0}{8 L} \sum_{n,m=1}^{N_M} U_{nm} \Delta N_{\rho + n} N_{\rho + m} \nonumber \\
&+& \sum_{q>0} \frac{v_0 q}{2} \sum_{n,m=1}^{N_M} U_{nm} 
\left\{b^\dagger_{\rho + n}(q) b_{\rho + m}(q) 
- \frac{1}{2} \left(b_{\rho + n}(q) b_{\rho + m}(q) 
+ b^\dagger_{\rho + m}(q) b^\dagger_{\rho + n}(q) \right) \right\}. 
\end{eqnarray}
Here $U_{nm} = 4 V_{nm}/(\pi v_0) = \left\{8 e^2/(\pi v_0
\kappa_{nm})\right\} \ln{(R_s/\bar R_{nm})}$ expresses the interaction
between the $n$-th and $m$-th shell with $\kappa_{nm}$  and $\bar
R_{nm}$
being the dielectric constant between the $n$-th and $m$-th shell and 
the ``mean radius''of $R_n$ and $R_m$ introduced in ref.\citen{Egger},
respectively.  
The modes except $(\rho +)$ are already diagonalized. 
By using the orthogonal matrix
satisfying $\sum_{m=1}^{N_M} U_{nm} \Gamma_{m,j} = \Gamma_{nj}g_j$, 
the Hamiltonian for $(\rho +)$ mode, ${\cal H}_{\rho +}$ is written as 
\begin{eqnarray}
 {\cal H}_{\rho +} &=& \frac{\pi v_0}{8L}\sum_{j=1}^{N_M} (1+g_j) (\Delta \tilde N_{\rho + j})^2 \nonumber \\
&+& \sum_{q>0}v_0 q\sum_{j=1}^{N_M} 
\left\{
(1+\frac{g_j}{2}) \tilde b^\dagger_{\rho + j}(q)\tilde b_{\rho + j}(q) 
- \frac{g_j}{4} ( \tilde b_{\rho + j}(q)\tilde b_{\rho + j}(q)  
+ \tilde b^\dagger_{\rho + j}(q)\tilde b^\dagger_{\rho + j}(q) )
\right\},   
\end{eqnarray}
where 
$\Delta N_{\rho + n} = \sum_{j=1}^{N_M} \Gamma_{nj} \Delta \tilde
N_{\rho + j}$,
$b_{\rho + n}(q) = \sum_{j=1}^{N_M} \Gamma_{nj} \tilde b_{\rho + j}(q)$
and $g_j$ is the eigenvalue of $U_{nm}$. 
Equation (47) is diagonalized by Bogoliubov transformation, 
$\tilde b_{\rho + j}(q) \to \cosh \varphi_j b_j(q)
- \sinh \varphi_j b_j^\dagger(q)$
($\e^{2 \varphi_j} = 1/\sqrt{1 + g_j} \equiv K_j$) as
\begin{eqnarray}
 {\cal H}_{\rho +} &=& \sum_{j=1}^{N_M} 
\left\{\frac{\pi v_{jN}}{8L}(\Delta \tilde N_{\rho + j})^2
+ \sum_{q>0} v_j q b^\dagger_j(q) b_j(q)\right\}, 
\end{eqnarray}
where $v_j = v_0/K_j$ and $v_{jN} = v_0 /K_j^2$ with $K_j = 1/\sqrt{1 +
g_j}$. 
In terms of $b_j(q)$ and $b^\dagger_j(q)$, $\phi_{\rho + n}$ is
expressed as 
\begin{eqnarray}
 \phi_{\rho + n}(x) &=& \sum_{j=1}^{N_M}\Gamma_{nj} 
\left\{
\cosh \varphi_j \phi_j(x) - \sinh \varphi_j \phi_j(-x)
\right\},  \\
\phi_j(x) &=& \sum_{q>0} \sqrt{\frac{\pi}{qL}}
\left\{
\e^{\im q x - \tilde a q/2} b_j(q) + h.c.
\right\}. 
\end{eqnarray}

\section{Charge distribution}
Based on the bosonization theory formulated above, 
the distribution of the charge density from the one boundary
is discussed with taking account of the shift 
of the chemical potential, $\mu$. 
The roles of the zero modes and of 
the bosonic long wavelength fluctuations are clarified.

At first, we consider the SWNT.
The term expressing the shift of the chemical potential, 
$-\mu \Delta N_{\rho +}$, is added to the Hamiltonian, eq.(40). 
Using eq.(21), the charge density is given as follows, 
\begin{eqnarray}
\langle :\sum_{s} a^\dagger_{p,s}(\bvec r) a_{p,s}(\bvec r): \rangle 
&=& \frac{a}{4N} \sum_{s,\alpha} 
\big\{
\langle \psi^\dagger_{R,\al,s}(x) \psi_{R,\al,s}(x) \rangle
+ \langle \psi^\dagger_{R,\al,s}(-x) \psi_{R,\al,s}(-x) \rangle \nonumber \\
& -& \e^{-\im 2 \al K_0 x}\langle \psi^\dagger_{R,\al,s}(x) \psi_{R,\al,s}(-x) \rangle
- \e^{\im 2 \al K_0 x}\langle \psi^\dagger_{R,\al,s}(-x) \psi_{R,\al,s}(x) \rangle
\big\}. 
\end{eqnarray}
The first and the second terms are spatially independent and given by 
\begin{eqnarray}
&&\sum_{s,\alpha} \left\{
\langle \psi^\dagger_{R,\al,s}(x) \psi_{R,\al,s}(x) \rangle 
+ \langle \psi^\dagger_{R,\al,s}(-x) \psi_{R,\al,s}(-x) \rangle   \right\} 
= \frac{\langle \Delta N_{\rho +} \rangle}{L}.  
\end{eqnarray}
On the other hand, 
the third and fourth terms are given by 
\begin{eqnarray}
&&\sum_{s,\alpha} \left\{ \e^{-\im 2 \al K_0 x}
\langle \psi^\dagger_{R,\al,s}(x) \psi_{R,\al,s}(-x) \rangle
+ \e^{\im 2 \al K_0 x}
\langle \psi^\dagger_{R,\al,s}(-x) \psi_{R,\al,s}(x) \rangle \right\} \nonumber \\ 
&=& 
\sum_{\al,s} 
\left\{
\langle \e^{-\im \frac{\pi x}{2 L}
\left\{\Delta N_{\rho +}+ \al (\Delta N_{\rho -} + 4 \nu / 3) 
   + s \Delta N_{\sigma +} + \alpha s \Delta N_{\sigma -} \right\}} \rangle 
\e^{-\im 2 \al K_0 x} \e^{\im f(2x)} + (x \to -x) 
\right\}  \nonumber \\
&\times& \frac{1}{2 \pi \tilde a} \left[\frac{\sinh^2(\pi \tia/2L)}{\sinh^2(\pi \tia/2L)+ \sin^2(\pi x/L)}\right]
^{(K_{\rho +}+3)/8} \nonumber \\
&\times& \prod_{l = 1}^{\infty}
\left[ 1 + \frac{\sin^2(\pi x/L)}{\sinh^2(l \pi^2 \xi_{\rho +}/L)}\right]
^{-K_{\rho +}/4} 
\left[ 1 + \frac{\sin^2(\pi x/L)}{\sinh^2(l \pi^2 \xi_0/L)}\right]
^{-3/4}, 
\end{eqnarray}
where $f(2x) = \tan^{-1}\left\{\sin(2\pi x/L)/(\e^{\pi \tia/L}-\cos2\pi
x/L)\right\}$, $\xi_{\rho +} = v_{\rho +}/(2 \pi T)$ and $\xi_0 = v_0/(2
\pi T)$. 
Here $\langle \cdots \rangle$ in the r.h.s. in eq. (53) expresses the average in terms of the zero
modes. 
Since the first line in the r.h.s in eq.(53) expresses the
oscillation of the charge density, 
the zero modes determine the uniform shift of the charge density by the
chemical potential, eq.(52), and the wavenumbers of the charge density
oscillation. 
On the other hand, the second and third lines are due to the
bosonic fluctuation and express the amplitude of the oscillation.    

We consider the semi-infinite case, $L \to \infty$. 
In this case, the average in terms of zero-modes 
is easily calculated,
\begin{eqnarray}
&& \langle \Delta N_{\rho +} \rangle = \frac{4 L \mu}{\pi v_{\rho + N}}, \\
&& \langle \e^{\pm\im \frac{\pi x}{2 L}
\left\{\Delta N_{\rho +}+ \al (\Delta N_{\rho -} + 4 \nu / 3) 
   + s \Delta N_{\sigma +} + \alpha s \Delta N_{\sigma -} \right\}} \rangle
= \e^{\pm \im 2 \mu x/v_{\rho + N}}, 
\end{eqnarray}
and $f(2x) = (\pi /2) {\rm sgn}(x)$. 
As a result, eq.(51) is calculated as follows
\begin{eqnarray}
\langle :\sum_s a^\dagger_{p,s}(\bvec r) a_{p,s}(\bvec r): \rangle 
&=& \frac{\mu a}{\pi N v_{\rho + N}} 
- 
\frac{a}{N}\frac{1}{2 \pi \tia} 
\left\{
\sin(2 K_0 + \frac{2\mu}{v_{\rho + N}})x - \sin(2 K_0 - \frac{2\mu}{v_{\rho + N}})x
\right\} A_S(x) \nonumber, \\ & & \\
A_S(x) &=& 
\left(\frac{\tia}{2x}\right)^{(K_{\rho +}+3)/4}
\left\{
\frac{\sinh(x/\xi_{\rho +})}{x/\xi_{\rho +}}
\right\}^{-K_{\rho +}/4}
\left\{
\frac{\sinh(x/\xi_0)}{x/\xi_0}
\right\}^{-3/4}.  
\end{eqnarray}
Note that eq.(56) together with eq.(57) in case of $T=0$ and $K_{\rho +} = 1$ 
reduces to the non-interacting case, eq.(6).
The deviation of the uniform charge density, $\mu a /(\pi N v_{\rho +
N}) = \mu a K_{\rho +}^2/(\pi v_0)$, and 
the shift of the wavenumber of the oscillation, $2 \mu / v_{\rho + N} =
2 \mu K_{\rho +}^2/v_0$,  
are smaller than those in the absence of the interaction.
Both quantities are about 0.04 as large as those in the absence of the
interaction for the typical SWNT. 
The fact means that 
Coulomb interaction prevents the carriers being injected
into the nanotube, and has the same origin as
suppression of the uniform charge susceptibility.     
The amplitude of the oscillation, $A_S(x)$, is large and decays slowly 
compared to the non-interacting case as is shown in Fig.3.
Here, the amplitude of the oscillation is shown as a function of
$x/\tia$ for $T/E_c = 0$ (solid curves) and $T/E_c = 0.04$ (dotted
curves) with $E_c = v_0/\tia$.
For each temperature,  the upper and lower curves express the amplitude 
in case of $K_{\rho +} =0.2$ and that in the absence of the
interaction, \ie $K_{\rho +} = 1$, respectively.  
The temperature, $T/E_c = 0.04$, corresponds to room temperature 
for $N=10$. 
Note that the amplitude delays as $\exp\left\{-(K_{\rho +}^2
+3)x/(4\xi_0)\right\}$ for $x \gg \xi_{\rho +}$. 
  \begin{figure}
\vspace{2em}
\centerline{\epsfxsize=7.0cm\epsfbox{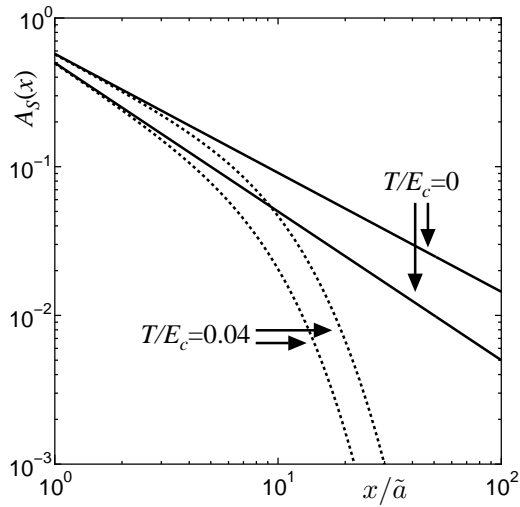}}
\caption{ The amplitude of Friedel oscillation of SWNT, $A_S(x)$, defined
   in eq.(57) as a function of $x/\tia$
   for $T/E_c = 0$ (solid curves) and $T/E_c = 0.04$ (dotted curves)
   where $E_c = v_0/\tia$.
For each temperature, 
the upper (lower) curve expresses the amplitude 
in case of $K_{\rho +} = 0.2$ ($K_{\rho +} = 1$). 
} 
    \label{fig:F-amp}
    \end{figure} 

Next we consider the case of the MWNT.
The charge density of the $n$-th shell is obtained for 
$L \to \infty$ as, 
\begin{eqnarray}
 & &\langle :\sum_s a^\dagger_{p,s,n}(\bvec r) a_{p,s,n}(\bvec r): \rangle 
\nonumber \\ 
&=& 
\frac{\mu a}{\pi N_n} \sum_{m,j = 1}^{N_M} \frac{\Gamma_{nj}\Gamma_{mj}}{v_{jN}} \nonumber \\
&-& \frac{a}{N_n} \frac{1}{2 \pi \tia} 
\left\{
\sin(2 K_0 + 2 \mu \sum_{m,j = 1}^{N_M} \frac{\Gamma_{nj}\Gamma_{mj}}{v_{jN}})x 
 - \sin(2 K_0 - 2 \mu \sum_{m,j = 1}^{N_M} \frac{\Gamma_{nj}\Gamma_{mj}}{v_{jN}}x)
\right\} \nonumber \\
&\times& \left(\frac{\tia}{2x}\right)^{(\sum_{j=1}^{N_M} \Gamma_{nj}^2 K_j +3)/4} 
\prod_{j=1}^{N_M}
\left\{
\frac{\sinh(x/\xi_{j})}{x/\xi_{j}}
\right\}^{-\Gamma_{nj}^2K_{j}/4}
\left\{
\frac{\sinh(x/\xi_0)}{x/\xi_0}
\right\}^{-3/4}  ,
\end{eqnarray}    
where $\xi_j = v_j/(2 \pi T)$. 
Because of the weak logarithmic dependence of $U_{nm}$ on 
${\bar R}_{nm}$, we can obtain sensible results by the approximation 
$U_{nm} = U$. 
The approximation leads to $g_1 = N_M U$, $g_j = 0$ for $j = 2 \cdots N_M$ and 
$\Gamma_{n1} = 1/\sqrt{N_M}$. 
As a result, the charge density
is obtained as 
\begin{eqnarray}
& &\langle :\sum_s a^\dagger_{p,s,n}(\bvec r) a_{p,s,n}(\bvec r): \rangle 
= 
\frac{\mu a}{\pi N_n v_{1N}} 
- \frac{a}{N_n} \frac{1}{2 \pi \tia} 
\left\{
\sin(2 K_0 + \frac{2 \mu}{v_{1N}})x - \sin(2 K_0 - \frac{2 \mu}{v_{1N}})x
\right\} A_M(x), \nonumber \\ & & \\
& & A_M(x) = \left(\frac{\tia}{2x}\right)^{(K_1 - 1)/(4N_M) + 1}  
\left\{
\frac{\sinh(x/\xi_{1})}{x/\xi_{1}}
\right\}^{-K_1/(4N_M)}
\left\{
\frac{\sinh(x/\xi_0)}{x/\xi_0}
\right\}^{-1 + 1/(4 N_M)}  .
\end{eqnarray}    
where $K_1 = 1/\sqrt{1 + N_M U}$, $\xi_1 = v_1/(2 \pi T)$ with $v_1 =
v_0/K_1$ and $v_{1N} = v_0/K_1^2$. 
The amplitude of the oscillation, $A_M(x)$, becomes small with increasing $N_M$
as is shown in Fig.3, where 
the quantity, $A_M(x)$, is shown as a function of $x/\tia$
   for $T/E_c = 0$ (solid curves) and $T/E_c = 0.04$ (dotted curves) 
with $\bar R_{nm} = 6$nm and $\kappa_{nm} = 1.4$ for several choices of $N_M$.
For each temperature, the curve from top to bottom corresponds to the
   case of $N_M = 1, 2, 5, \infty$. 
We note that the effects of the interaction on $A_M(x)$ disappear
in the limit, $N_M \to \infty$.  
  \begin{figure}
\vspace{2em}
\centerline{\epsfxsize=7.0cm\epsfbox{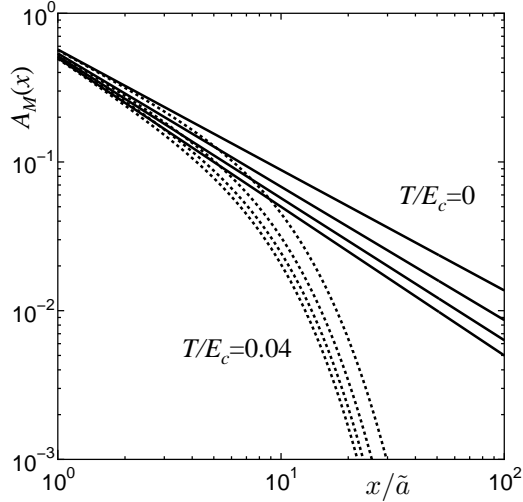}}
\caption{ The amplitude of Friedel oscillation of MWNT as a function of $x/\tia$
   for $T/E_c = 0$ (solid curves) and $T/E_c = 0.04$ (dotted curves) 
with $\bar R_{nm} = 6$nm and $\kappa_{nm} = 1.4$ for several
   choices of $N_M$.   
For each temperature, the curve from top to bottom corresponds to the
   case of $N_M = 1, 2, 5, \infty$.  
} 
    \label{fig:M-F-amp}
    \end{figure}

\section{Summary and discussion}

In the present paper, the bosonization theory
with open boundaries was developed for both SWNT and MWNT.
We payed attention to the relationship between 
the electron operator of the tight-binding model 
and the slowly varying field.   
Based on this bosonization theory, 
we investigated the distribution of the charge density from an edge 
with taking account of the shift of the chemical potential expressing 
carrier doping to the nanotube. 

The bosonized Hamiltonian is written by the sum of
symmetric/antisymmetric combination of the charge/spin excitation. 
Each mode has two contribution, zero modes and bosonic
fluctuation. 
Effects of the Coulomb interaction with the strongest amplitude 
appear in only the symmetric charge excitation. 
These facts are the same as the case of the periodic boundary
condition. 
However,    
the Hamiltonian of zero modes are expressed in terms of only the extra number.
This fact is due to the open boundary condition, and 
different from the periodic boundary condition, where 
the zero mode Hamiltonian is written by the extra number and the current.    
The charge distribution in the presence of the shift of the chemical
potential has two component. 
One is spatially independent and the other shows oscillatory behavior 
which vanishes in the absence of the chemical potential.  
The zero modes determine the magnitude of the spatially independent 
component and the wavenumbers of the oscillation. 
The long-range Coulomb interaction strongly suppress the both quantities.  
This means the fact that the interaction prevents carriers from being
injected into the nanotubes. 
For the typical SWNT with $K_{\rho +} = 0.2$, 
the both quantities are about 0.04 as large as those in the absence of
the interaction. 
The quantity, the shift of the chemical potential, 
corresponds to the applied gate voltage or difference between the work function 
of the nanotube and that of the substrate material. 
The carrier doping from the gold substrate has been observed by 
scanning tunneling spectroscopy (STS)\cite{Wildoer,Venema} 
and scanning tunneling microscopy (STM)\cite{Venema-II}.
In the STS experiments, the asymmetry of the density of states has been 
observed, and it is maintained to be due to the difference between the
work function of the nanotube and that of the gold(111) substrate. 
In the present formalism, the shift of the density of states 
is just given by the chemical potential even in the presence of 
the Coulomb interaction,\cite{Okamura} which is the same as the conclusion in
refs.\citen{Wildoer} and \citen{Venema}. 
However, the deviation of the Fermi wavenumber, $\delta k$, in the presence of 
the chemical potential is given by not the simple form of $\mu/v_0$ 
but $\mu/v_{\rho + N} = \mu K_{\rho +}^2/v_0$ as is seen in eq.(56). 
On the other hand, the STM experiment observes $\delta k \sim
\mu/v_0$. 
The discrepancy seems to need the further theoretical study on the Coulomb
interaction of the nanotube. 
The amplitude of the oscillation determined by 
the bosonic long wavelength fluctuation shows the non-integer power law
behavior as a function of the spatial coordinate at $T=0$. 
We found that the amplitude is enhanced by the
interaction and it becomes smaller with increasing the number of the
metallic shells, $N_M$, in the MWNT and identical with that in the
absence of the interaction in the case, $N_M \to \infty$. 
Finally, we note that the present result of the charge distribution 
are also valid for the strong impurity potential.\cite{Fabrizio,Egger-Grabert,EG-II}     
 
\section*{Acknowledgment}
The authors would like to thank S. Iwabuchi and C. Tanaka for 
variable discussion. 
This work was supported by a Grant-in-Aid for Encouragement of 
Young Scientist (No. 13740220) and  
Grant-in-Aid for Scientific Research (A) (No. 13304026) and (C)
(No. 14540302) 
from the Ministry of Education, Culture, Sports, Science and Technology,
Japan.


\end{document}